\documentclass[a4paper,11pt]{article}
\pdfoutput=1 % if your are submitting a pdflatex (i.e. if you have
\usepackage{jinstpub} % for details on the use of the package, please
\usepackage[T1]{fontenc} 
\usepackage[utf8]{inputenc}

\title{Decay tunnel instrumentation for the ENUBET neutrino beam}

\author[a,b]{F. Acerbi}
\author[e,p]{A. Berra}
\author[e]{M. Bonesini}
\author[c,d]{A. Branca}
\author[e,f]{C. Brizzolari}
\author[c]{G. Brunetti}
\author[r]{M. Calviani}
\author[d,g]{S. Carturan}
\author[e,p]{S. Capelli}
\author[h]{M.G. Catanesi}
\author[r]{N. Charitonidis}
\author[i]{S. Cecchini}
\author[i]{F. Cindolo}
\author[c,d]{G. Collazuol}
\author[c]{E. Conti}
\author[c]{F. Dal Corso}
\author[j,k]{G. De Rosa}
\author[c,d]{C. Delogu}
\author[e,f]{A. Falcone}
\author[a]{A. Gola}
\author[l]{C. Jollet}
\author[r]{V. Kain}
\author[m]{B. Klićek}
\author[n]{Y. Kudenko}
\author[c,d]{M. Laveder}
\author[c,d]{A. Longhin}
\author[o]{L. Ludovici}
\author[e,p]{E. Lutsenko}
\author[h,q]{L. Magaletti}
\author[i]{G. Mandrioli}
\author[i]{A. Margotti}
\author[e,p,1]{V. Mascagna\note{Corresponding author.}}
\author[i]{N. Mauri}
\author[e,f]{L. Meazza}
\author[l]{A. Meregaglia}
\author[c]{M. Mezzetto}
\author[r]{M. Nessi}
\author[c,d]{M. Pari}
\author[t]{A. Paoloni}
\author[e,f]{E. Parozzi}
\author[i,s]{L. Pasqualini}
\author[a]{G. Paternoster}
\author[i]{L. Patrizii}
\author[i]{M. Pozzato}
\author[e,p]{M. Prest}
\author[c]{F. Pupilli}
\author[h]{E. Radicioni}
\author[j,k]{C. Riccio}
\author[j,k]{A.C. Ruggeri}
\author[c,d]{C. Scian}
\author[i]{G. Sirri}
\author[m]{M. Stipćevic}
\author[i]{M. Tenti}
\author[e,f]{F. Terranova}
\author[e,f]{M. Torti}
\author[e]{E. Vallazza}
\author[r]{F. Velotti}
\author[c,d]{M. Vesco}
\author[t]{L. Votano}

\affiliation[a]{Fondazione Bruno Kessler (FBK),  Via Sommarive 18 - 38123 Povo (TN), IT}
\affiliation[b]{INFN-TIFPA, Università di Trento, Via Sommarive 14 - 38123 Povo (TN), IT}
\affiliation[c]{INFN Sezione di Padova, via Marzolo 8 - 35131 Padova, IT}
\affiliation[d]{Università di Padova, via Marzolo 8 - 35131 Padova, IT}
\affiliation[e]{INFN Sezione di Milano-Bicocca, Piazza della Scienza 3 - 20133 Milano, IT}
\affiliation[f]{Università di Milano-Bicocca, Piazza della Scienza 3 - 20133 Milano, IT}
\affiliation[g]{INFN, Laboratori Nazionali di Legnaro, Viale dell'Università 2 - 35020 Legnaro (PD), IT}
\affiliation[h]{INFN Sezione di Bari, Via Giovanni Amendola 173 - 70126 Bari, IT}
\affiliation[i]{INFN Sezione di Bologna, viale Berti-Pichat 6/2 - 40127 Bologna, IT}
\affiliation[j]{INFN, Sezione di Napoli, Strada Comunale Cinthia - 80126 Napoli, IT}
\affiliation[k]{Università ``Federico II'' di Napoli, Corso Umberto I 40 - 80138 Napoli, IT}
\affiliation[l]{CENBG, Universitè de Bordeaux, CNRS/IN2P3, 33175 Gradignan, FR}
\affiliation[m]{Center of Excellence for Advanced Materials and Sensing Devices, Ruđer Bošković
Institute, HR-10000 Zagreb, HR}
\affiliation[n]{Institute of Nuclear Research of the Russian Academy of Science, 142190 Moscow, RU}
\affiliation[o]{INFN Sezione di Roma 1, Piazzale A. Moro 2, 00185 Rome, IT}
\affiliation[p]{Università degli Studi dell'Insubria, Via Valleggio 11 - 22100 Como, IT}
\affiliation[q]{Università degli Studi di Bari, Via Giovanni Amendola 173 - 70126 Bari, IT}
\affiliation[r]{CERN, Esplanade des particules - 1211 Genève 23, CH}
\affiliation[s]{Università degli Studi di Bologna, viale Berti-Pichat 6/2 - 40127 Bologna, IT}
\affiliation[t]{INFN, Laboratori Nazionali di Frascati, via Fermi 40 - 00044 Frascati (Rome), Italy}

\emailAdd{valerio.mascagna@uninsubria.it}

\abstract{The uncertainty in the initial neutrino flux is the main limitation for a precise determination of the absolute neutrino cross section. The ERC funded ENUBET project (2016-2021) is studying a facility based on a narrow band beam to produce an intense source of electron neutrinos with a ten-fold improvement in accuracy. Since March 2019 ENUBET is also a Neutrino Platform experiment at CERN: NP06/ENUBET. A key element of the project is the instrumentation of the decay tunnel to monitor large angle positrons produced together with $\nu_e$ in the three body decays of kaons ($K_{e3}$) and to discriminate them from neutral and charged pions. The need for an efficient and high purity e/$\pi$ separation over a length of several meters, and the requirements for fast response and radiation hardness imposed by the harsh beam environment, suggested the implementation of a longitudinally segmented Fe/scintillator calorimeter with a readout based on WLS fibers and SiPM detectors. An extensive experimental program through several test beam campaigns at the CERN-PS T9 beam line has been pursued on calorimeter prototypes, both with a shashlik and a lateral readout configuration. The latter, in which fibers collect the light from the side of the scintillator tiles, allows to place the light sensors away from the core of the calorimeter, thus reducing possible irradiation damages with respect to the shashlik design.
This contribution will present the achievements of the prototyping activities carried out, together with irradiation tests made on the Silicon Photo-Multipliers.
The results achieved so far pin down the technology of choice for the construction of the 3 m long demonstrator that will take data in 2021.

}

\proceeding{15$^{\text{th}}$ Topical Seminar on Innovative Particle and Radiation Detectors\\
  14-17 October 2019\\
  Siena, Italy}

\begin{document}

\maketitle
\flushbottom

\section{The ENUBET project}
\label{sec:the_enubet_project}
\begin{figure}[htb!]
    \centering
    \includegraphics[width=0.9\textwidth]{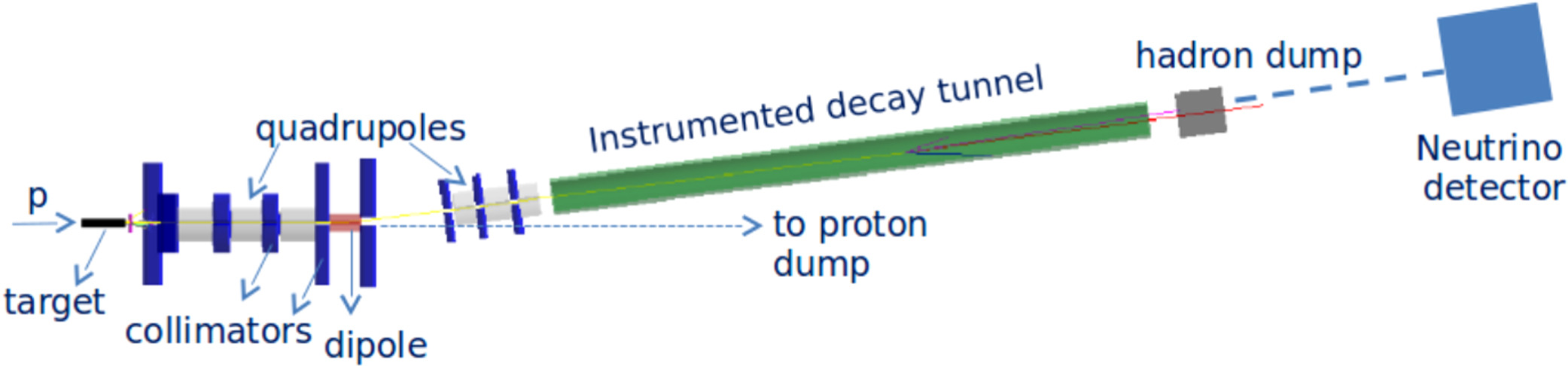}
    \caption{The ENUBET beamline.}
    \label{fig:beamline}
\end{figure}

Modern neutrino cross section measurements performed at accelerators are limited by the knowledge on the initial $\nu$ flux which is, in turn, affected by uncertainties on the hadro-production and the overall particle propagation along the decay tunnel. The ENUBET project aims to demonstrate that the so far achieved precision (5\%--10\%) can be improved by one order of magnitude by a direct measurement of the electron neutrinos generated in an instrumented tunnel where the positrons produced in kaon decays (K$\rightarrow e^+ \pi^0 \nu_e$) are detected and tagged~\cite{eoi}. The project includes the design of a conventional narrow-band beam in a facility consisting of a short ($\sim$20~m) transfer line followed by a 40~m long decay tunnel. After the protons hit a target, the produced particles are focused, momentum selected and transported towards the decay tunnel entrance. The hadron selection and the tunnel length are optimized so that the only source of neutrinos is the three body semi-leptonic decay of kaons (K$_{e3}$)~\cite{epjc2015, spsc2018}. In the proposed beamline (Fig.~\ref{fig:beamline}), the electron neutrino flux is monitored by detecting the large-angle emitted positrons in the decay tunnel using longitudinally segmented shashlik calorimeters. They are sampling calorimeters in which the light is collected by Wavelength Shifters (WLS) fibers that cross perpendicularly the tiles of both absorbing and scintillating material. This technology is well established and cost effective. A good energy resolution is achievable and adjustable by choosing a convenient absorbing/scintillating tiles thickness and fiber density. The baseline option foresees shashlik modules of 4.3~$X_0$ thickness with a compact light readout through a matrix of 3$\times$3 WLS ($\sim$one fiber/cm$^2$) directly coupled to small-area SiPMs hosted on a PCB on the back of each module, hence avoiding dead zones from fiber bundling. The neutrino tunnel instrumentation will be complemented by a photon veto system to tag positrons from $K_{e3}$ decays and suppress the gamma background from $\pi_0$ decays.

\section{The positron tagger}
The ENUBET tagger consists in a modular calorimeter for e/$\pi$ separation readout by WLS fibers and Silicon PhotoMultipliers. Rings of plastic scintillators below the calorimeter (``t$_0$-layer'') allow to reject photons from $\pi^0$ decays . During the first 3 years of the collaboration, an intense prototyping activity has been performed including tests on beamlines and radiation campaigns for the photosensors.  
\subsection{Shashlik prototype with integrated readout}\label{sec:shashlik}
\begin{figure}[htb!]
    \centering
    \includegraphics[width=0.7\textwidth]{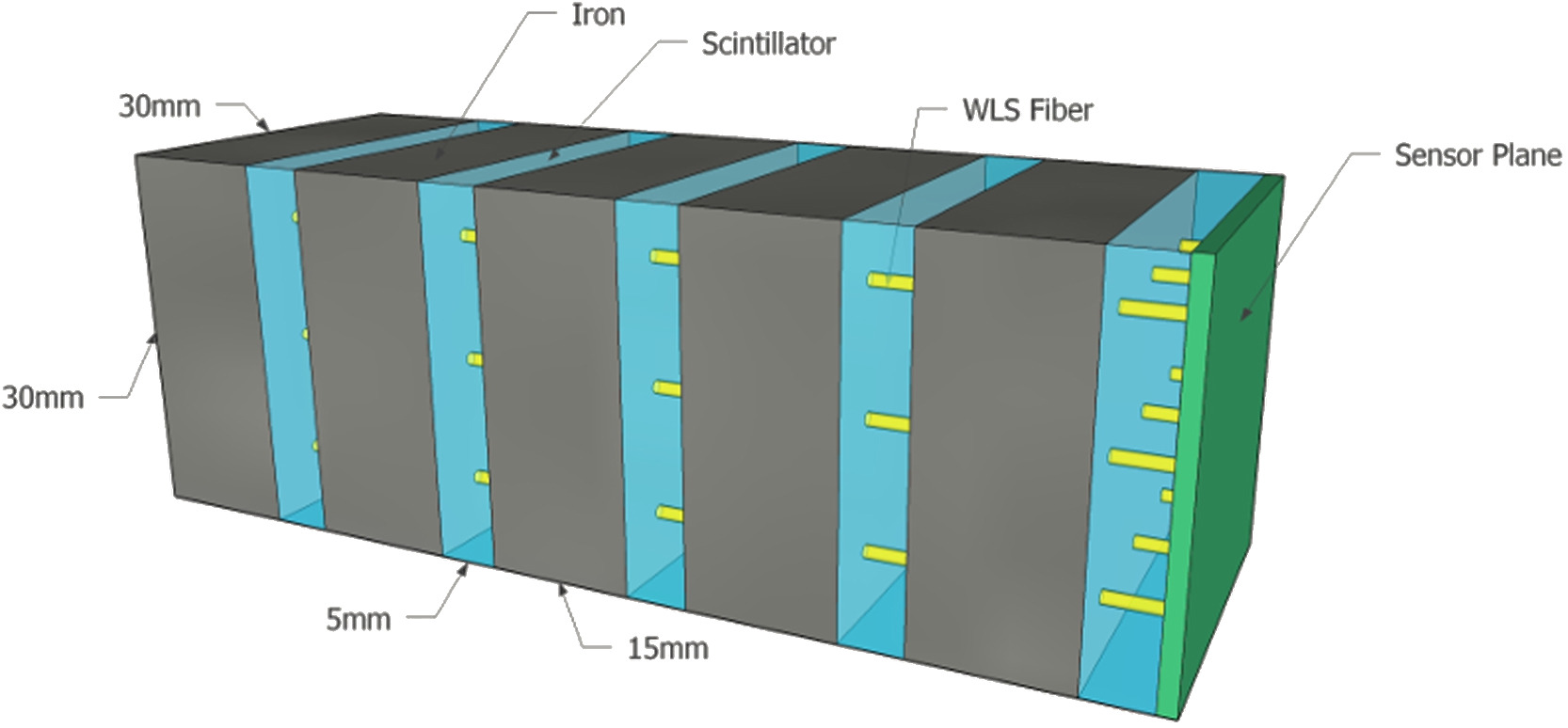}
    \caption{Scheme of a UCM with the embedded readout board. The length is 10~cm (4.3~$X_0$) and the cross section 3$\times$3 cm$^2$. The scintillation light is collected by 9 WLS fibers each one coupled to a SiPM hosted on the PCB readout board.}
    \label{fig:ucm}
\end{figure}

\begin{figure}[htb!]
    \centering
    \includegraphics[width=\textwidth]{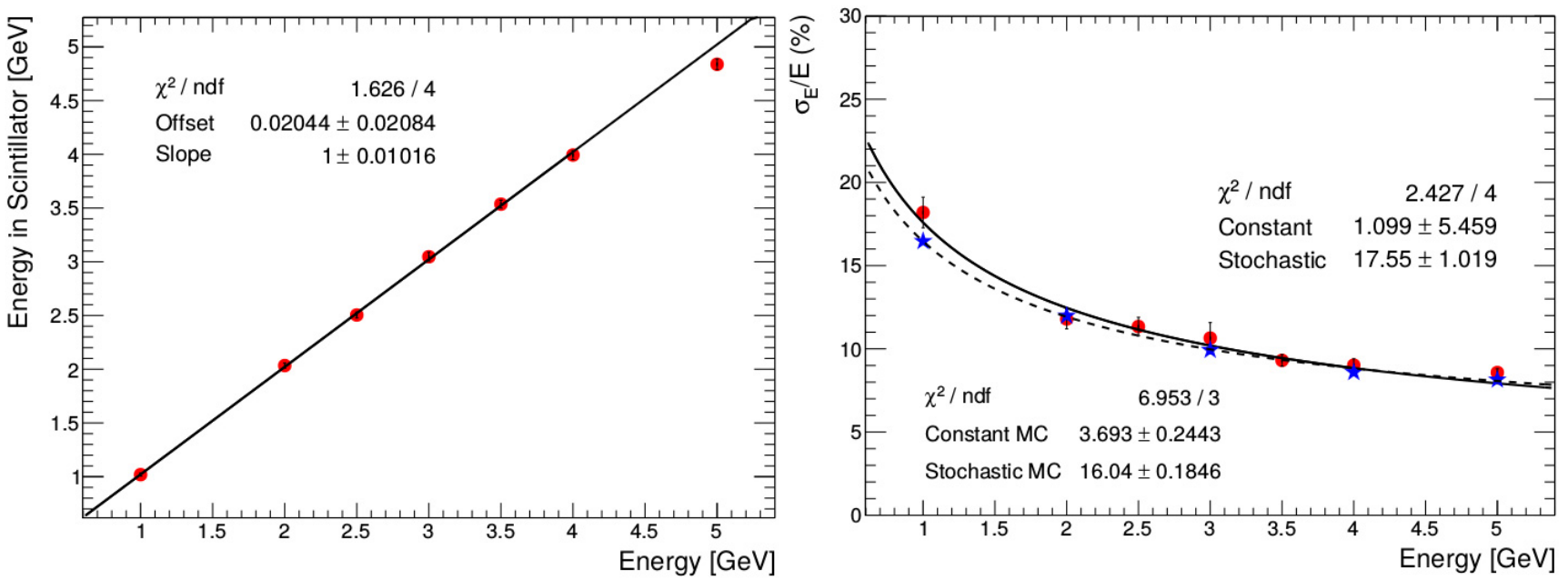}
    \caption{Linearity (left) and energy resolution (right) obtained with the first prototype tests in 2016~\cite{ieee2017}.}
    \label{fig:linearity_and_resolution}
\end{figure}

The first prototypes consisted in fundamental units called Ultra Compact Modules (UCMs, Fig.~\ref{fig:ucm}) equipped with embedded SiPMs in the calorimeter bulk~\cite{nim2016, ieee2017, nim2019}. 
Each UCM with a transverse size of 3$\times$3~cm$^2$ (1.7 Moli\`ere radii), consists in alternated tiles of iron (15~mm thickness) and plastic scintillator (5~mm thickness), for a total length of 10~cm corresponding to 4.3~radiation lengths (X$_0$). The scintillator tiles have been painted with Titanium dioxide (TiO$_2$) diffusive coating (EJ-510) to increase the light collection efficiency and they have been drilled with a Computer Numerical Control (CNC) machine for the 9 holes hosting the WLS fibers. Two types of plastic scintillators -- WLS fiber combinations have been tested: EJ-200 with Kuraray Y-11 fibers and BC-412 with Bicron BCF-92. Having the first one a higher light yield it was used in the majority of the prototypes, while for the t$_0$-layer tests the second one (faster response) seemed to be preferable. One prototype with a structure of 3$\times$2$\times$2 UCMs has been tested in June 2016~\cite{ieee2017} at the T9 beamline of the CERN PS East Area facility~\cite{t9}. The prototype has been exposed to electrons, muons, and pions in the energy range of interest for neutrino physics applications (1–5 GeV). The test resulted in a linear response up to 4~GeV (Fig.~\ref{fig:linearity_and_resolution}, left) and an energy resolution of 18\% (11\%) at 1~GeV (4~GeV) as show on the rigth of Fig.~\ref{fig:linearity_and_resolution}.
Both quantities are in agreement with the specification obtained by MC simulations for positron tagging. A bigger prototype made of 7$\times$4$\times$2 UCMs  corresponding to 30 X$_0$ was tested in November of the same year confirming the previous results~\cite{jinst2018}.

\subsection{Polysiloxane scintillator prototype}
\begin{figure}[htb!]
    \centering
    \includegraphics[width=\textwidth]{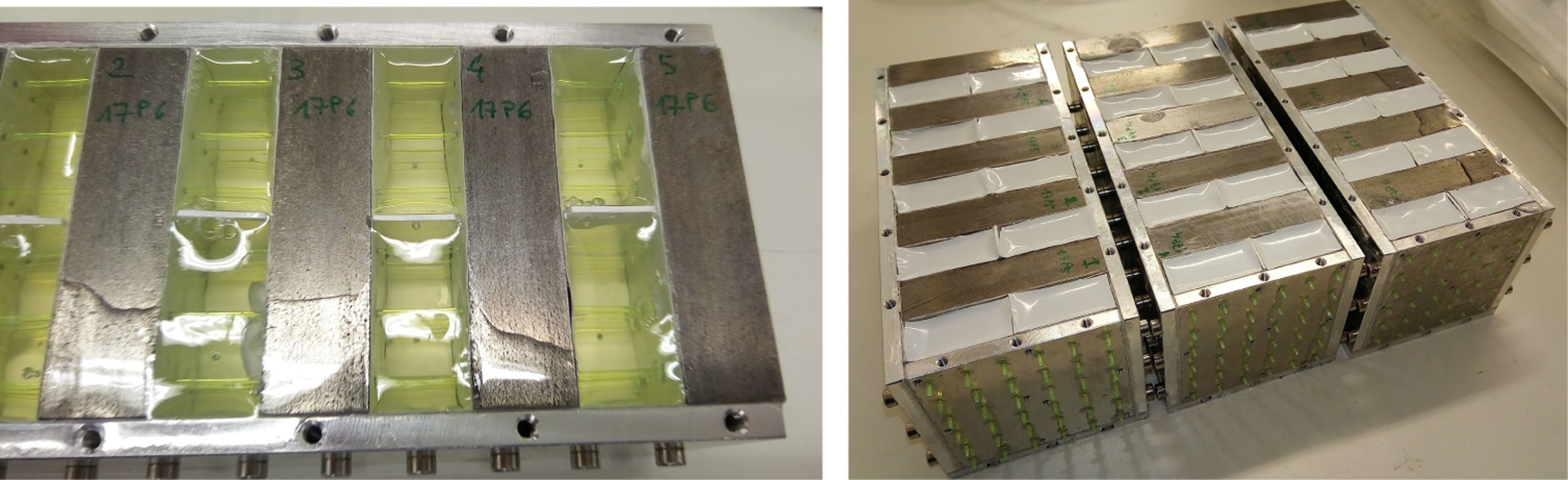}
    \caption{Pictures of the polysiloxane prototype before (left) and after (right) wrapping the scintillator with a Tyvek film. The prototype had a size of 6$\times$6$\times$45~cm$^3$ resulting in a depth of 13~X$_0$.}
    \label{fig:polysiloxane}
\end{figure}

During the prototyping phase, a silicon-based scintillator has been used as an alternative to the standard plastic scintillator. Polysiloxane based materials present a larger radiation tolerance, a reduced ageing process and the transparency of the used solution is preserved up to a 10 kGy exposure~\cite{poly2010, poly2011}. In the manufacturing procedure, suitable dyes are dissolved in polysiloxane in the liquid phase, before reticulation, similarly to what is done for plastic scintillators. The solution is then poured in the mechanical case hosting the fibers and the iron absorber tiles from the top and then baked in an oven. The polysiloxane eventually reticulates acquiring a rubber-like texture. This technique ensures a good coupling to the fibers avoiding the mechanical stress caused by the drilling process in rigid materials. The drawback is the reduced light yield (one third of the EJ200 per unit of length). To increase the light yield a new set of UCMs has been built with 15~mm of active material (instead of 5~mm as in the case of plastic scintillator) and a prototype array of 3$\times$2$\times$2 UCMs has been assembled (Fig.~\ref{fig:polysiloxane}) and tested at the end of 2017 at the CERN T9 beamline.
The test resulted in a good linearity (within ~3\% in the 1--5~GeV energy range), a performance in the $e/\pi$ separation comparable with the standard scintillator prototype and an energy resolution of 17\% at 1~GeV~\cite{nima2020}. 

\subsection{Lateral readout scheme}\label{sec:lateral}
\begin{figure}[htb!]
    \centering
    \includegraphics[width=\textwidth]{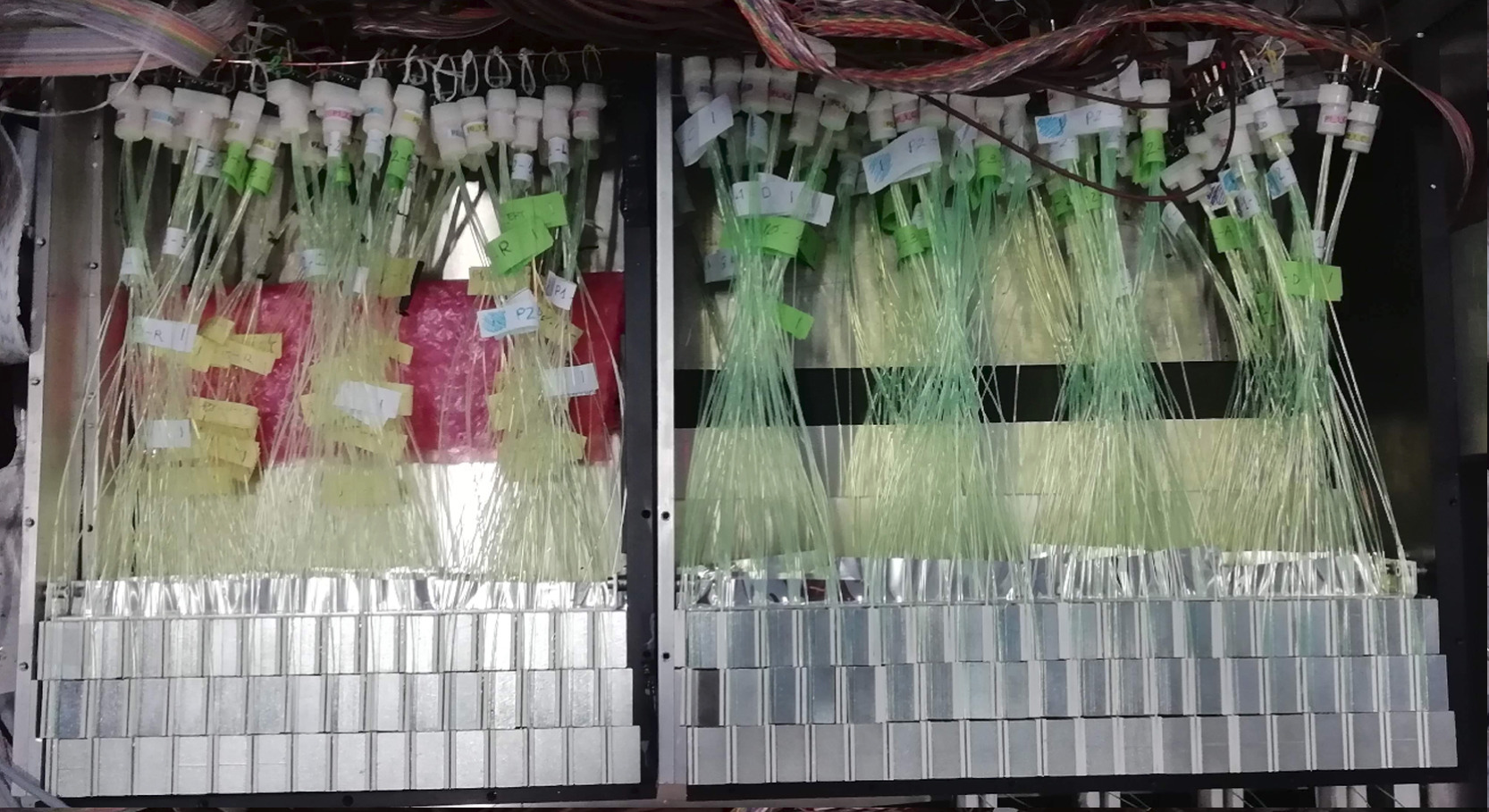}
    \caption{Side view of the prototype with the lateral readout scheme built in 2018. The array of UCMs is on the bottom while SiPMs for the readout of the WLS fibers are on the top at an average distance of $\sim$30~cm from the bulk.}
    \label{fig:lateral}
\end{figure}

During 2018 a new design has been developed: the light is collected from both
sides of the scintillator tile through WLS fibers arranged in a lateral groove (10 fibers per UCM are bundled and optically coupled to a single SiPM). The SiPMs for
the readout are placed outside of the calorimeter bulk at a distance of around 30~cm (Fig.~\ref{fig:lateral}). 
This design is less compact, but on the other hand reduces the possible neutron damage, and prevents the exposure of Silicon to hadronic showers. Neither the absorber nor the scintillator have to be drilled anymore but grooved on the sides, which is a safer (and cheaper) machine work. Moreover, this scheme allows a faster maintenance and/or replacements of the readout sensors. In 2018 two prototypes have been tested at the CERN East Hall facility with a e-$\mu$-$\pi$ beam in the 1--5~GeV range: a smaller one made of 3$\times$2$\times$2 UCMs (each alternating 15~mm of iron and 5~mm of EJ-204 plastic scintillator) and a bigger prototype consisting in a 7$\times$4$\times$3 structure of the same UCMs. Preliminary results show that the linearity and energy resolution are comparable with the ones obtained with the shashlik readout scheme~\cite{jinst2020}. MC simulations are in agreement with the collected data and the analysis performed so far. Since the lateral readout calorimeter shows performance similar to the shashlik detector but has no significant irradiation damage, it has been chosen as the preferred option for ENUBET. A large size demonstrator of the ENUBET instrumented tunnel, based
on this technology will be built in 2021.   

\section{Radiation tests on the SiPMs}
\begin{figure}[htb!]
    \centering
    \includegraphics[width=\textwidth]{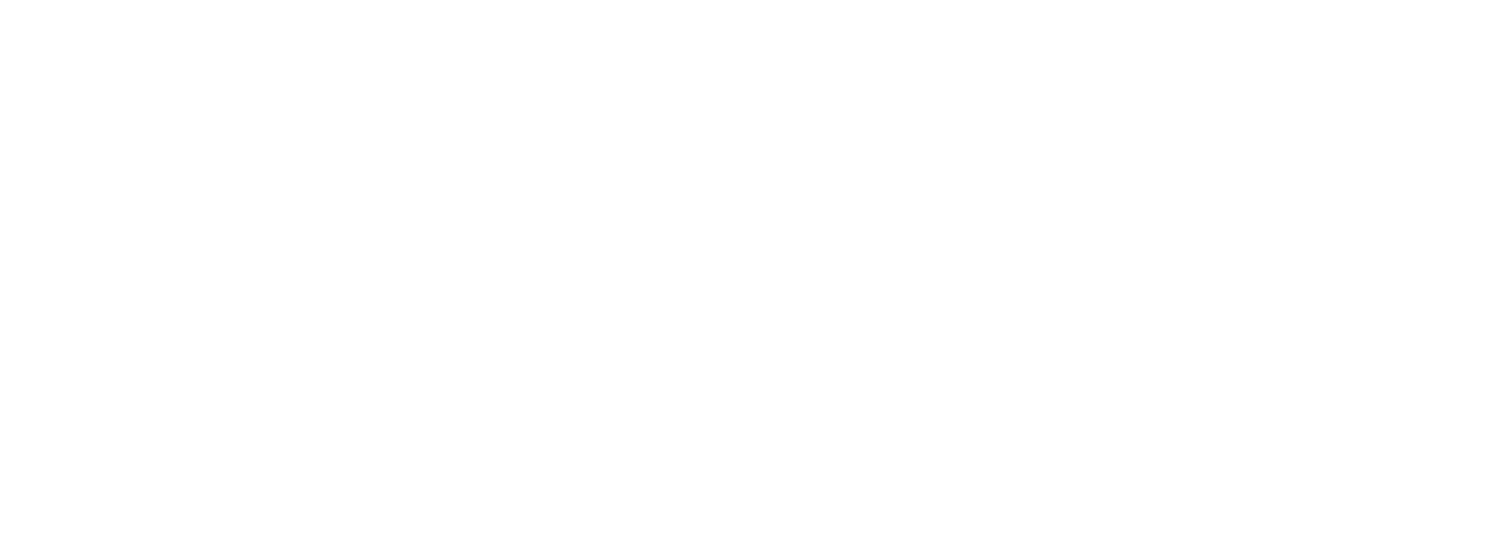}
    \caption{Test beam results obtained before (left) and after (right) SiPM irradiation. The black line represents the calorimeter energy spectrum. The red line excludes charged pions from the results, while the green line excludes the muons (MIPs) as well. The fit line on the electron peak, obtained with this dose absorption, demonstrates good capabilities of the calorimeter in e/$\pi$ separation.}
    \label{fig:irrad}
\end{figure}

\begin{figure}[htb!]
    \centering
    \includegraphics[width=0.8\textwidth]{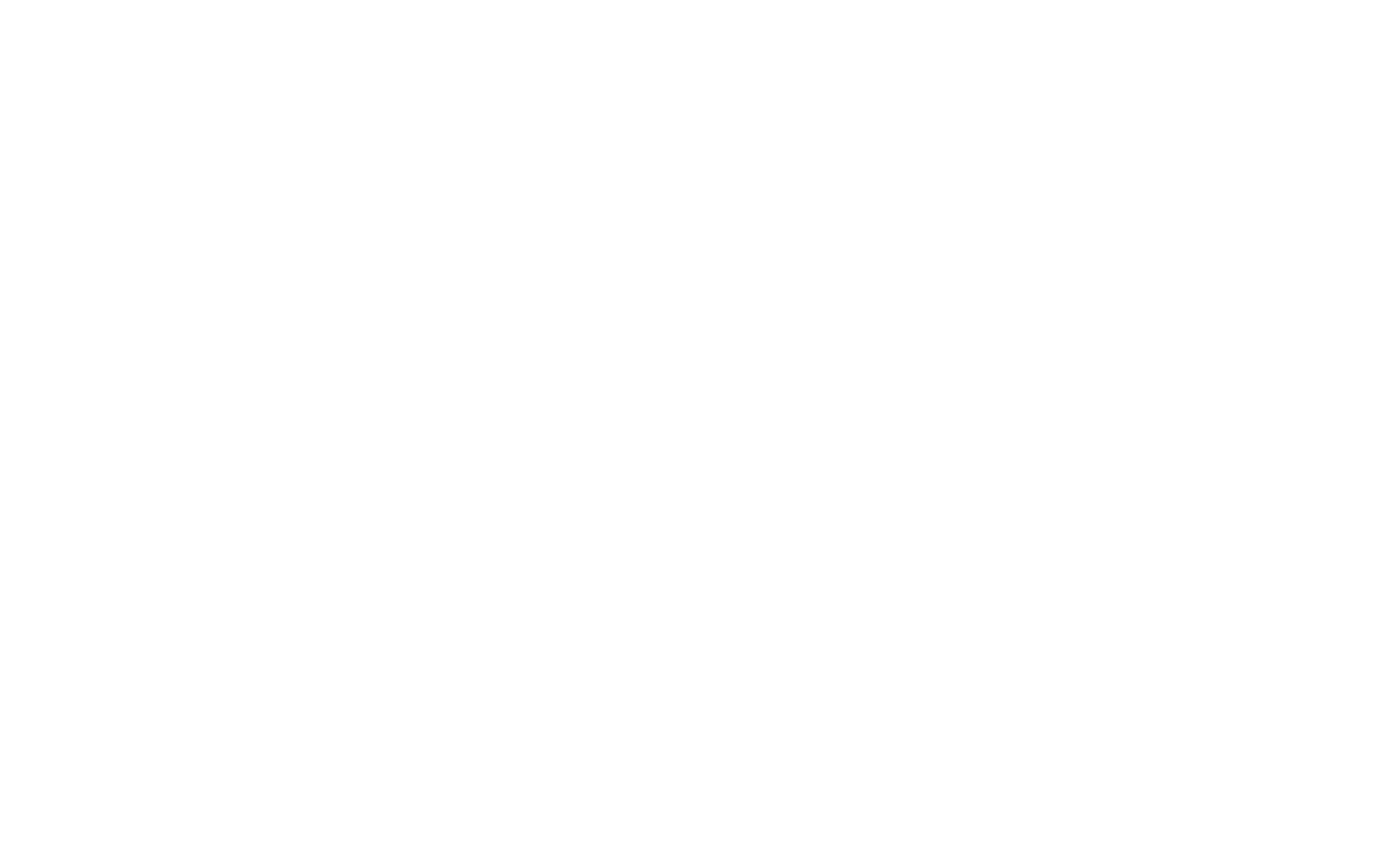}
    \caption{Beam energy spectrum obtained choosing a 13.5~mm thick scintillator instead of 5~mm. The MIP signal ($\sim$10~mV) enhanced in the red curve, is still well distinguished from the noise pedestal ($\sim$5~mV). Same color code of Fig.~\ref{fig:irrad}}
    \label{fig:irrad_more_scintillator}
\end{figure}

In the lateral readout scheme (Sec.~\ref{sec:lateral}), the expected neutron flux on the SiPMs is lower than the original shashlik design (Sec.~\ref{sec:shashlik}). Nevertheless a radiation campaign has been performed at the CN facility at INFN-LNL (Legnaro, Italy)~\cite{jinst2019}. In the worst case and for a 5-year experiment lifetime, the expected neutron fluence is of the order of 10$^{11}$ 1-MeV-eq-n/cm$^2$. SiPM RGB-HD devices developed by FBK\footnote{https://www.fbk.eu/en/}) with different pixel size (12-15-20~$\mu$m) have been exposed to an integrated fluence of 10$^{12}$ n/cm$^2$ using 5 MeV protons from the CN Van der Graaf generator impinging on a thick Beryllium target. The sensors have been fully characterized during the irradiation and after several months from it. It has been demonstrated that the sensitivity to a single photo-electron is maintained up to an integrated dose of 3$\times$10$^9$~1-MeV-eq-n/cm$^2$. During the tests performed at the CERN East Hall T9 beamline described in the previous sections, the neutron irradiated sensors have been used to compare the results with the ones before the irradiation. The test showed that with a 10$^{11}$~n/cm$^2$ dose, the calorimeter performances of the detector are not compromised: the electron-MIP peaks ratio is conserved (Fig.~\ref{fig:irrad}) and a good separation between the MIP peak and the background pedestal is still possible increasing the scintillator thickness (Fig.~\ref{fig:irrad_more_scintillator}). Moreover, the gain reduction due to the absorbed radiation can be compensated increasing the overvoltage within the SiPM operational range. 

\section{Conclusions}

An extensive experimental program has been carried out in order to design and test detectors suitable to instrument the decay tunnel of the monitored neutrino beam proposed by the ENUBET project.
Prototypes of a segmented iron-scintillator calorimeter for e/$\pi$ separation and of a system for $\gamma$ rejection based on plastic scintillator tiles read out by WLS coupled to SiPMs have been fully characterized.
In particular the lateral readout of the scintillator tiles of the calorimeter modules had proven to be the preferred solution for the ENUBET scope, since in this layout possible irradiation damages to the SiPMs are effectively reduced.
The ENUBET demonstrator, that will take data in 2021, will be built exploiting this technology.

Furthermore we successfully operated for the first time in high energy physics a calorimeter with polysiloxane-based scintillators. It could represent an intriguing alternative for applications that need to stand sustained irradiation levels.

\acknowledgments
This project has received funding from the European Unions Horizon 2020 Research and Innovation program under Grant Agreement no.~\href{https://cordis.europa.eu/project/id/681647}{681647}.

% We suggest to always provide author, title and journal data:
% in short all the informations that clearly identify a document.

\end{document}